\documentclass[epj]{webofc}
\usepackage[utf8]{inputenc}
\usepackage[varg]{txfonts}   % Web of Conferences font
\usepackage{booktabs}
\usepackage{xcolor}
\definecolor{darkred}{rgb}{0.4,0.0,0.0}
\definecolor{darkgreen}{rgb}{0.0,0.4,0.0}
\definecolor{darkblue}{rgb}{0.0,0.0,0.4}
\usepackage[bookmarks,linktocpage,colorlinks,
    linkcolor = darkred,
    urlcolor  = darkblue,
    citecolor = darkgreen]{hyperref}
\usepackage{subfigure}
\wocname{EPJ Web of Conferences}
\woctitle{Lattice2017}
\begin{document}
%%%%%%%%%%%%%%%%%%%%%%%%%%%%%%%%%%%%%%%%%%%%%%%%%%%%%%%%%%%%%%%%%%%%%%%%%%%%%
%
\selectlanguage{english}
%----------------------------------------------------------------------------
\title{%
Pion mass dependence of the HVP contribution to muon \begin{boldmath}
$\!g-2$ \end{boldmath}
}
%----------------------------------------------------------------------------
\author{%
\firstname{Maarten} \lastname{Golterman}\inst{1}
\fnsep\thanks{Speaker, \email{maarten@sfsu.edu}} \and
\firstname{Kim} \lastname{Maltman}\inst{2,3} \and
\firstname{Santiago}  \lastname{Peris}\inst{4}
% etc.
}
%----------------------------------------------------------------------------
\institute{%
Department of Physics and Astronomy, San Francisco State University,
San Francisco, CA 94132, USA
\and
Department of Mathematics and Statistics,
York University, Toronto, ON Canada M3J~1P3
\and
CSSM, University of Adelaide, Adelaide, SA~5005 Australia
\and
Dept. of Physics and IFAE-BIST, Univ. Aut\`onoma de Barcelona, E-08193 Bellaterra, Barcelona, Spain
}
%----------------------------------------------------------------------------
\abstract{%
  One of the systematic errors in some of the current lattice computations of the HVP contribution to the muon anomalous magnetic moment $g-2$ is that associated with the extrapolation to the physical pion mass. We investigate this extrapolation assuming lattice pion masses in the range of 220 to 440 MeV with the help of two-loop chiral perturbation theory, and find that such an extrapolation is unlikely to lead to control of this systematic error at the 1\% level. This remains true even if various proposed tricks to improve the chiral extrapolation are taken into account.
}
%----------------------------------------------------------------------------
\maketitle
%----------------------------------------------------------------------------
\section{Introduction}\label{intro}

The leading-order (LO), hadronic vacuum polarization (HVP) contribution to the muon anomalous
magnetic moment $a_\mu=(g-2)/2$ is given by the integral
\begin{equation}
\label{amu}
a_\mu^{\rm HLO}=-4\alpha^2\int_0^\infty\frac{dQ^2}{Q^2}\,w(Q^2)\left(
\Pi(Q^2)-\Pi(0)\right)\ ,
\end{equation}
where $\Pi(Q^2)$ is obtained from the electromagnetic (EM) hadronic vacuum polarization
\begin{equation}
\label{HVP}
\Pi_{\mu\nu}(Q)=\left(\delta_{\mu\nu}Q^2-Q_\mu Q_\nu\right)\Pi(Q^2)\ ,
\end{equation}
$\alpha$ is the fine-structure constant, and $w(Q^2)$ is a known weight function
\cite{ER,TB2003}.   The integrand in Eq.~(\ref{amu}) is strongly peaked at 
$Q^2\sim m_\mu^2/4\sim (50~\mbox{MeV})^2$ because of the presence of the
factor $w(Q^2)/Q^2$.   With the smallest momentum
on a periodic volume at present being at least 100~MeV or larger, all the usual
systematic errors afflicting a precision determination of $\Pi(Q^2)$ are ``magnified,''
and it is thus important to investigate each of the sources of systematic error in detail,
especially if the aim is a computation of $a_\mu^{\rm HLO}$ with an accuracy better
than 1\%.   Here, we consider the extrapolation to the physical pion mass from
a lattice computation using pion masses in the range 220 to 440~MeV, assuming 
that all other errors (in particular, lattice spacing and finite volume effects) 
would be under control.   In particular, we investigate the use of the so-called
``ETMC'' trick \cite{Fengetal}, and variants thereof, which were designed to improve the 
chiral behavior of $a_\mu^{\rm HLO}$, {\it i.e.}, to ``flatten'' the dependence of
$a_\mu^{\rm HLO}$ on $m_\pi$.

For the published version of this work, see Ref.~\cite{Golterman:2017njs}.

%----------------------------------------------------------------------------
\section{The ETMC trick}\label{ETMC}

In Ref.~\cite{Fengetal}, it was proposed to use, instead of Eq.~(\ref{amu}), the expression\footnote{We discuss a specific example of the ETMC trick, the one which has been most used in
practice.}
\begin{equation}
\label{amuETMC}
a_\mu^{\rm HLO}=-4\alpha^2\int_0^\infty\frac{dQ^2}{Q^2}\,w(Q^2)\left(
\Pi\left(\frac{m^2_{\rho,{\rm latt}}}{m^2_\rho}Q^2\right)-\Pi(0)\right)\ ,
\end{equation}
where $m_\rho=775$~MeV is the physical rho mass, while $m_{\rho,{\rm latt}}$ is the
rho mass extracted from the lattice ensemble also used to compute $\Pi(Q^2)$.   Of course, 
at the physical pion mass (assuming extrapolation to the continuum limit and infinite 
volume), we have that $m_{\rho,{\rm latt}}=m_\rho$, and Eq.~(\ref{amuETMC}) is 
identical to Eq.~(\ref{amu}).  

The observation that the extrapolation to the physical pion mass using Eq.~(\ref{amuETMC})
may be flatter than that using Eq.~(\ref{amu}) rests on the observation that vector-meson
dominance (VMD) gives a fairly good approximation to the HVP.   Inspired by VMD, let us assume
a model for the HVP of the form
\begin{equation}
\label{VMD}
\Pi^{\rm VMD}(Q^2)=\frac{f_\rho^2 m_\rho^2}{m_\rho^2+Q^2}+\Delta\Pi(Q^2)\ ,
\end{equation}
where $\Delta\Pi(Q^2)$ estimates the difference between perturbation theory and VMD
for $Q^2\gg m_\rho^2$.   It follows that, with $\Pi(Q^2)$ computed on the lattice
\begin{equation}
\label{VMDlatt}
\Pi^{\rm VMD}\left(\frac{m^2_{\rho,{\rm latt}}}{m^2_\rho}Q^2\right)=\frac{f_\rho^2 m^2_{\rho,{\rm latt}}}{m^2_{\rho,{\rm latt}}+\frac{m^2_{\rho,{\rm latt}}}{m^2_\rho}Q^2}+\Delta\Pi\left(\frac{m^2_{\rho,{\rm latt}}}{m^2_\rho}Q^2\right)=\frac{f_\rho^2 m_\rho^2}{m_\rho^2+Q^2}+\Delta\Pi\left(\frac{m^2_{\rho,{\rm latt}}}{m^2_\rho}Q^2\right)\ ,
\end{equation}
where we assumed that the dependence of $f_{\rho,{\rm latt}}\approx f_\rho$, for which there is
some evidence.   As the perturbative tail for large $Q^2$, $\Delta\Pi(Q^2)$,  only
contributes a few percent to the total $a_\mu^{\rm HLO}$, we see that, to the extent that VMD
gives a good estimate, indeed Eq.~(\ref{amuETMC}) is less dependent on $m_\pi$ than
Eq.~(\ref{amu}).

This simple argument suggests that indeed using Eq.~(\ref{amuETMC}) instead of Eq.~(\ref{amu})
should improve the extrapolation to the physical pion mass.   However, it is well known that the
two-pion cut also contributes significantly (of order 10\%) to $a_\mu^{\rm HLO}$, and a detailed
investigation is necessary to check the improvement in accuracy that can be gained with this trick.
In fact, in Ref.~\cite{Fengetal}, the lowest pion mass in the lattice computations was about 
300~MeV, and the results of that work suggest that the extrapolation from that value to the
physical value is rather long.   A variant of this trick was also used by HPQCD in Ref.~\cite{HPQCD}, where
the lowest-order pion loop contribution ({\it cf.} Sec.~\ref{ChPT} below) at the lattice pion mass was subtracted first, before applying the ETMC trick to the modified data, after which the lowest-order
pion loop was added back in at the physical pion mass.   We will also consider this variant in the
investigation we report on below.   We will do so using next-to-next-to-leading-order (NNLO)
Chiral Perturbation Theory (ChPT) for the vacuum polarization.

%----------------------------------------------------------------------------
\section{Chiral perturbation theory}\label{ChPT}

To simplify matters, we will first consider the $I=1$ component of the HVP, returning later to the
full EM case.
The results we will need from ChPT were calculated in Ref.~\cite{ABT}.  For $I=1$, schematically, it looks like
\begin{equation}
\label{chpt}
\Pi^{I=1}(Q^2)-\Pi^{I=1}(0)=-4(F(Q^2)-F(0))-\frac{4Q^2}{f_\pi^2}F^2(Q^2)+\frac{16Q^2}{f_\pi^2}
L_9F(Q^2)+8C_{93}Q^2+C(Q^2)^2\ ,
\end{equation}
where $F(Q^2)$ is a known function (representing two-pion and two-kaon cuts), and
$L_9=0.00593(43)$ \cite{BT}, $C_{93}=-0.01536(44)$~GeV$^{-2}$ \cite{C93}
and $C=0.289$~GeV$^{-4}$
\cite{C93} are NLO, NNLO, and NNNLO low-energy constants (LECs), respectively.
$C$ represents the analytic $(Q^2)^2$ dependence at NNNLO, but only models the
$Q^2$ dependence at that order, because the non-analytic terms at that order are not known.

\begin{figure}[t] % no figure before 1st section
  \centering
  \includegraphics[width=10cm,clip]{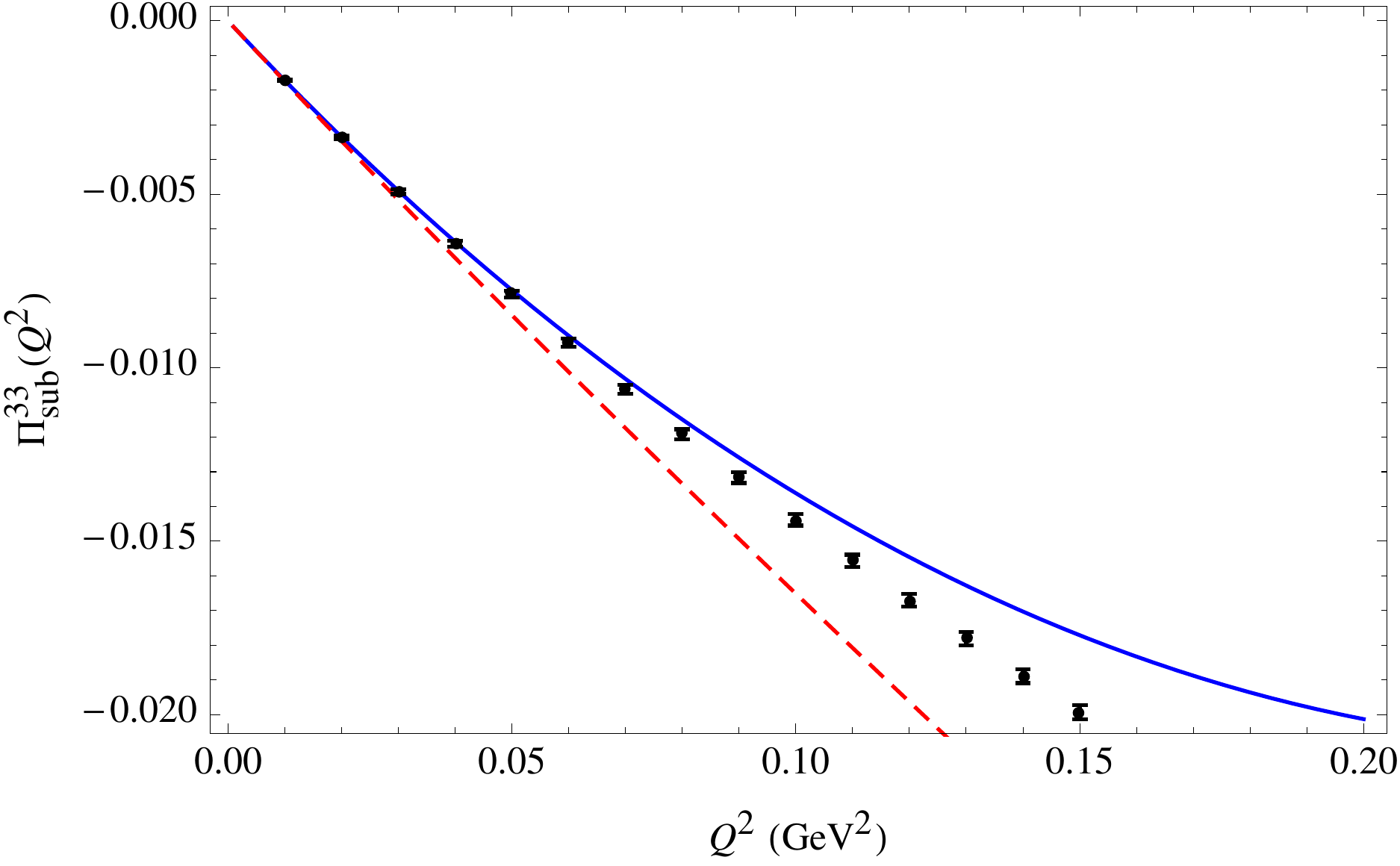}
  \caption{Comparison of the HVP between data from hadronic tau decays (black points)
  and ChPT.   The blue solid curve includes the $C$ term in Eq.~(\ref{chpt}) with value
  $C'=0.289$~GeV$^{-4}$; the red dashed curve has $C=0$.}
  \label{fig-1}% Give a unique label
\end{figure}

Figure~\ref{fig-1} shows the comparison between ChPT and data for $\Pi^{I=1}(Q^2)-\Pi^{I=1}(0)$,
where the data were extracted from the ALEPH data for non-strange, vector hadronic tau
decays \cite{ALEPH13}.   The construction of these data points from the data, using fits to 
extend the measured spectral function beyond $m_\tau^2$ \cite{alphas14}, was described in
detail in Ref.~\cite{Golterman:2017njs}.   The ChPT curves in the figure include the $C$ term
in Eq.~(\ref{chpt}) (blue solid curve), or have it set equal to zero (red dashed curve).   The figure
shows that ChPT agrees well with the data to about $Q^2\sim 0.1$~GeV$^2$, if the 
phenomenological $C$ term is included.   In order to make this quantitative, we consider
a cut-off version of $a_\mu^{\rm HLO}$:
\begin{equation}
\label{amucutoff}
a_\mu^{{\rm HLO},I=1}(Q^2_{max})=\left(\frac{\alpha}{\pi}\right)^2\int_0^{Q^2_{\max}}\frac{dQ^2}{Q^2}\,w(Q^2)\left(
\Pi^{I=1}(Q^2)-\Pi^{I=1}(0)\right)\ ,
\end{equation}
so that for $Q^2_{max}$ small enough, we can use the ChPT representation for the integrand.
Taking $Q^2_{max}=0.1$~GeV$^2$, we find that
\begin{eqnarray}
\label{comp}
\hspace{3cm}a_\mu^{{\rm HLO},I=1}(0.1)&=&\ \ 9.81\times 10^{-8}\ ,\qquad\mbox{(data)}\ ,\\
&=&\ \ 9.73\times 10^{-8}\ ,\qquad\mbox{(ChPT)}\ ,\nonumber\\
a_\mu^{{\rm HLO},I=1}(\infty)&=&11.95\times 10^{-8}\ ,\qquad\mbox{(data)}\ .\nonumber
\end{eqnarray}
This shows that for $Q^2_{max}=0.1$~GeV$^2$ ChPT agrees with the data to better than
1\%, and that the interval between 0 and $0.1$~GeV$^2$ yields 82\% of the value of
$a_\mu^{{\rm HLO},I=1}$.   

We will use $a_\mu^{{\rm HLO},I=1}(0.1)$ in order to investigate the dependence on the
pion mass, using ChPT in order to do so.   As already discussed in Sec.~\ref{ETMC}, also
the rho mass depends on the pion mass, and it is this dependence which lies at the heart
of the trick employed in Refs.~\cite{Fengetal,HPQCD} to extrapolate to the physical pion
mass.  In ChPT, dependence on the rho mass will show up through the mass dependence of effective versions, $C_{93,{\rm latt}}^{\rm eff}$ and $C_{\rm latt}^{\rm eff}$, of the 
(mass-independent) LECs $C_{93}$ 
and $C$  obtained in fits employing the form~(\ref{chpt}) above while also incorporating higher-order 
mass-dependent terms in the chiral expansion.  Assuming VMD, one expects these effective 
LECs to scale with $m_\rho$ as 
\begin{equation}
\label{LECeff}
C_{93,{\rm latt}}^{\rm eff} = \hat{C}_{93}^{\rm eff}\frac{m_\rho^2}{m_{\rho,{\rm latt}}^2}\ ,\qquad
C_{\rm latt}^{\rm eff} = \hat{C}\frac{m_\rho^4}{m_{\rho,{\rm latt}}^4}\ , 
\end{equation}
where $\hat{C}_{93}^{\rm eff}$ and $\hat{C}$ are the effective LECs at the physical pion mass. We will employ Eq.~(\ref{chpt}) with $C_{93}$ and $C$ replaced by
$C_{93,{\rm latt}}^{\rm eff}$ and $C_{\rm latt}^{\rm eff}$ to construct a model for use in calculating $a_\mu^{{\rm HLO},I=1}(Q^2_{max})$ for
non-physical values of the pion mass and exploring the impact of the ETMC and HPQCD tricks.

%----------------------------------------------------------------------------
\begin{boldmath}
\section{Pion mass dependence of $a_\mu^{{\rm HLO}}(0.1)$}\label{piondep}
\end{boldmath}
We will first consider the pion mass dependence of $a_\mu^{{\rm HLO},I=1}$ by considering the
contribution up to $Q^2_{max}=0.1$~GeV$^2$, {\it i.e.}, $a_\mu^{{\rm HLO},I=1}(0.1)$.
Of course, the ``missing'' part (the contribution from $Q^2>0.1$~GeV$^2$) also depends 
on the pion mass.   However, the pion mass dependence of that part is inherently different
from that of $a_\mu^{{\rm HLO},I=1}(0.1)$, and it would be a fluke if this would change the
lessons we will learn from considering $a_\mu^{{\rm HLO},I=1}(0.1)$.   Moreover, we find that
these lessons do not change when we take $Q^2_{max}=0.2$~GeV$^2$ \cite{Golterman:2017njs}.

In order to study the pion mass dependence of $a_\mu^{{\rm HLO},I=1}(0.1)$, we need the 
pion mass dependence of $m_K$ and $f_\pi$, both of which appear in Eq.~(\ref{chpt}), as
well as of $m_\rho$, which appears ``indirectly,'' through Eq.~(\ref{LECeff}).   As an example,
we take a range of pion masses between 220~MeV and 440~MeV, with corresponding
values for $m_K$, $f_\pi$ and $m_\rho$ from MILC \cite{MILC}.
They are shown in Table~\ref{tab-1}. 
\begin{table}[thb]
  \small
  \centering
  \caption{Masses and $f_\pi$ values used in our tests.  All numbers in MeV.}
  \label{tab-1}% Give a unique label
  \begin{tabular}{c|c|c|c}\toprule
  $m_\pi$  & $m_K$ & $f_\pi$ & $m_\rho$  \\\midrule
  223 & 514 & 98 & 826 \\
  262 & 523 & 101 & 836 \\
  313 & 537 & 104 & 859 \\
  382 & 558 & 109 & 894 \\
  440 & 581 & 114 & 929 \\  \bottomrule
  \end{tabular}
\end{table}
%%%%%%%%%%%%%%%%%%%
\begin{figure}[t]
\vspace*{4ex}
\centering
\includegraphics*[width=6cm]{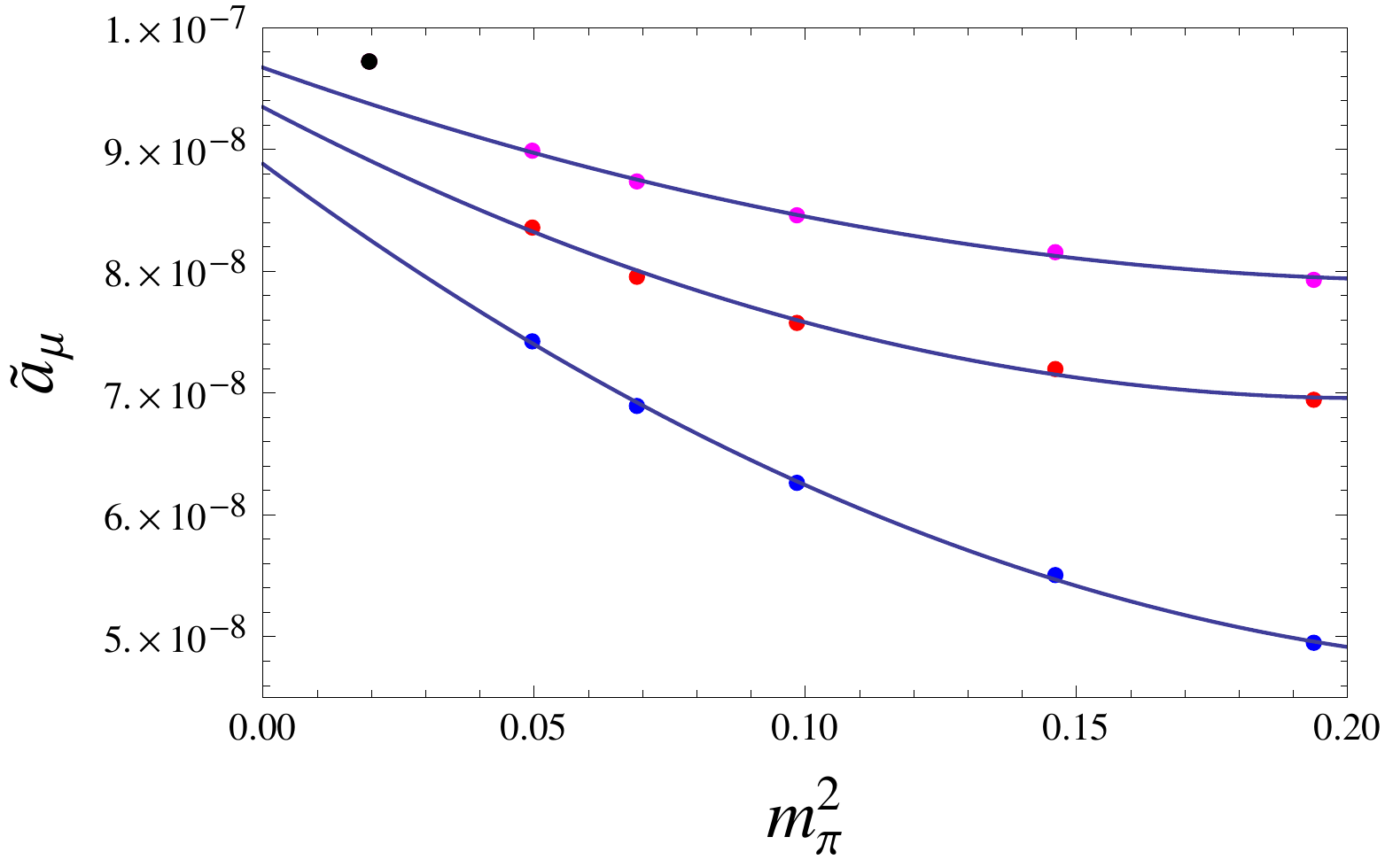}
\hspace{1ex}
\includegraphics*[width=6cm]{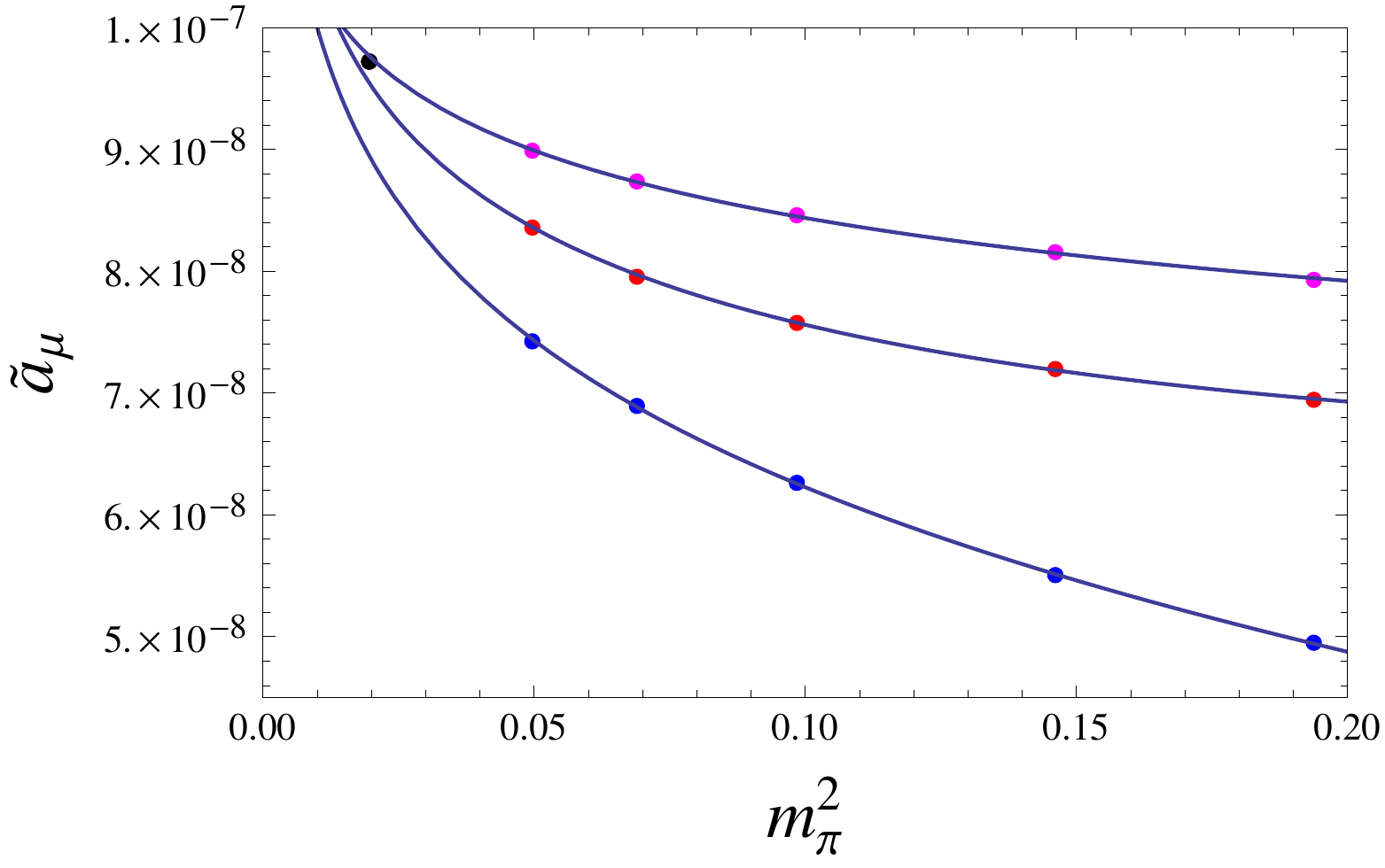}\\
\vspace*{4ex}
\includegraphics*[width=6cm]{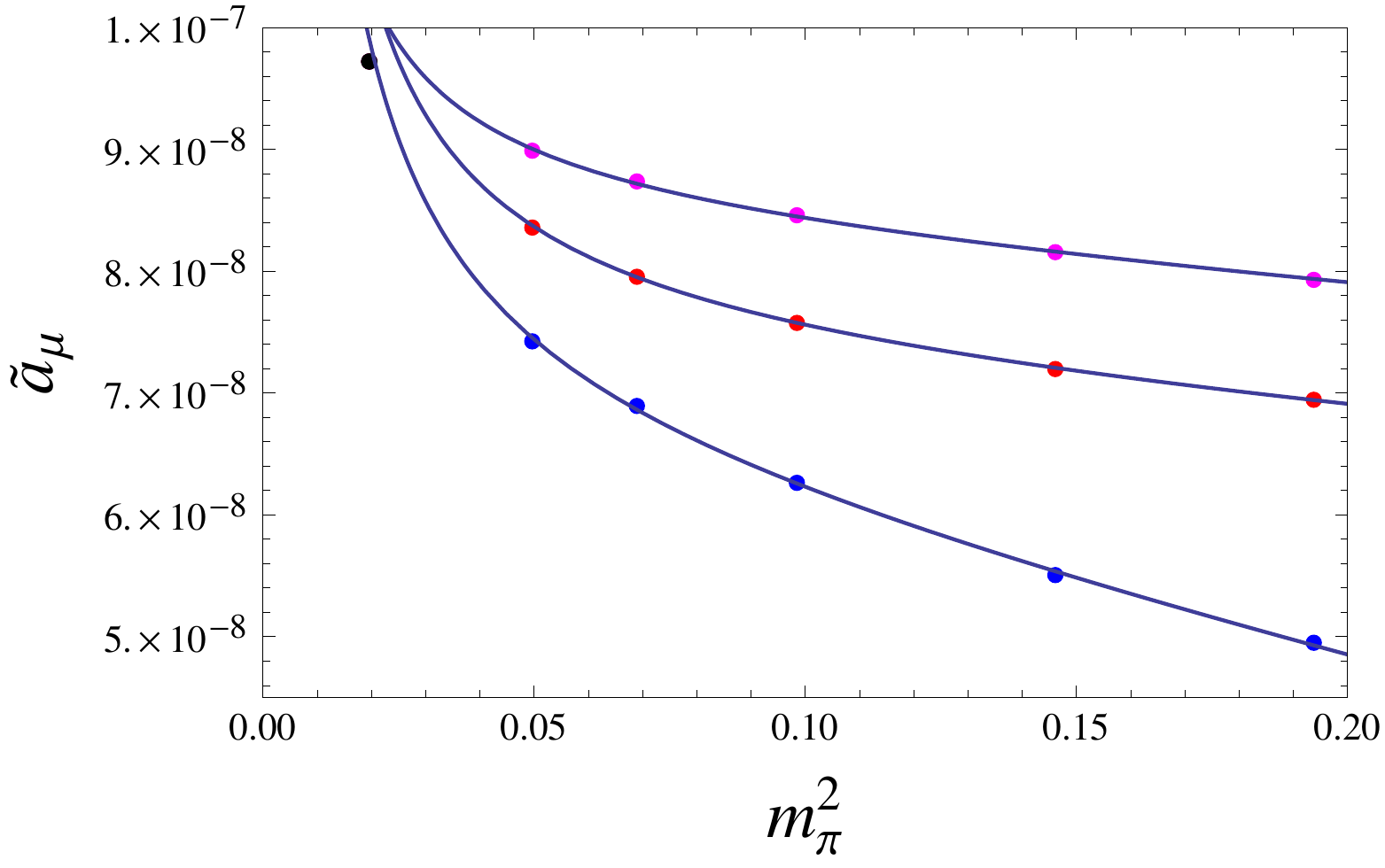}
\caption%
{The unmodified and ETMC- and HPQCD-improved versions of 
$\tilde{a}_\mu\equiv a_\mu^{{\rm HLO},I=1}(0.1)$ as a function of $m_\pi^2$. In each plot, the upper (magenta) data 
points are HPQCD-improved, the middle (red) data points ETMC-improved
and the lower (blue) data points unimproved.
The (black) point in the upper left corner of each plot is the 
``physical'' point, $a_\mu^{{\rm HLO},I=1}(0.1)=9.73\times 10^{-8}$ ({\it cf.} Eq.~(\ref{comp})).
Fits are ``quadratic'' (upper left panel), ``log'' (upper right panel)
and ``inverse'' (lower panel). For further explanation, see text.}
\label{fig-2}
\vspace*{-4ex}
\end{figure}
%%%%%%%%%%%%%%%%%%%
We use these values to calculate, using ChPT, three values of $a_\mu^{{\rm HLO},I=1}(0.1)$, and
show these in Fig.~\ref{fig-2}, for each of the lines in Table~\ref{tab-1}.   The black dot shows the value at the physical point; while the
points at the larger pion masses show the values at the points of Table~\ref{tab-1}, 
with the lower set corresponding to the unmodified values for  $a_\mu^{{\rm HLO},I=1}(0.1)$,
the middle set the values modified by the use of the ETMC trick, and the upper set of points
those modified by the HPQCD version of the ETMC trick.

The three different panels all show the same points, but the fits are different.   The curves
correspond to fits to the functional forms
\begin{eqnarray}
\label{fits}
a_\mu^{{\rm HLO},I=1,{\rm quad}}(0.1)&=&Am_{\pi,{\rm latt}}^4+Bm_{\pi,{\rm latt}}^2+C\ ,\qquad\qquad\qquad\!\!\mbox{(quadratic})\ ,\\
a_\mu^{{\rm HLO},I=1,{\rm log}}(0.1)&=&A\log(m_{\pi,{\rm latt}}^2/m_\pi^2)+Bm_{\pi,{\rm latt}}^2+C\ ,\qquad\mbox{(logarithmic})\ ,\nonumber\\
a_\mu^{{\rm HLO},I=1,{\rm in}}(0.1)&=&A/m_{\pi,{\rm latt}}^2+Bm_{\pi,{\rm latt}}^2+C\ ,\qquad\qquad\quad\ \mbox{(inverse})\ ,\nonumber
\end{eqnarray}
with the upper left panel showing the quadratic fit, the upper right panel showing the logarithmic
fit, and the lower panel the inverse fit.   The logarithmic form is inspired by ChPT, even though
the values of $m_\pi$ for which the chiral logarithm should be visible are far smaller than the
physical pion mass (see below).   The inverse
fit was employed in Ref.~\cite{HPQCD}.

The figures show that all fits lead to extrapolations that miss the physical point, with possible
exceptions for the logarithmic fit if one uses either version of the ETMC-modified points, and
the inverse fit if one uses the unmodified points.   We quantify these deviations in Table~\ref{tab-2}.
\begin{table}[thb]
  \small
  \centering
  \caption{Values for $a_\mu^{{\rm HLO},I=1}(0.1)\times 10^8$ for the three types of
fit ({\it cf.} Eq.~(\ref{fits})) and the three data sets. For reference, the correct
model value is $a_\mu^{{\rm HLO},I=1}(0.1)\times 10^8=9.73$ ({\it cf.} Eq.~(\ref{comp})).}
  \label{tab-2}% Give a unique label
  \begin{tabular}{c|c|c|c}\toprule
    & unimproved data & ETMC-improved data  & HPQCD-improved data  \\\midrule
  quadratic fit & 8.26 & 8.91 & 9.38 \\
  logarithmic fit & 8.96 & 9.55 & 9.77 \\
  inverse fit & 9.93 & 10.46 & 10.33 \\  \bottomrule
  \end{tabular}
\end{table}
In the case of real QCD, we will not have access to the ``exact'' value of $a_\mu^{{\rm HLO},I=1}(0.1)$, as we have here in this test, of course.    Therefore, one will have to resort to a different
method for estimating the systematic error associated with the extrapolation to the physical
pion mass.   With no solid theoretical guidance on what the fit form should be, the best option is
to take the difference between two ``reasonable'' fits.   Taking, for instance, the relative difference
between extrapolations using the quadratic and logarithmic fits, one finds differences of 
8\%, 7\% and 4\% for the unimproved, ETMC-improved and HPQCD-improved data.   We see
that indeed the two variants of the ETMC trick do improve the chiral extrapolation, but not 
sufficiently to reach the desired accuracy of below 1\%.   We have also redone this exercise
omitting the largest pion mass (440~MeV), but found that this makes no significant difference.
We emphasize that the logarithmic fit is {\it not} theoretically preferred.   While ChPT does predict a
logarithm as in Eq.~(\ref{fits}) \cite{Golterman:2017njs}, its coefficient is far smaller than the values we find in unconstrained
fits.   The reason is that the scale for the chiral logarithm as predicted in ChPT is set by the muon
mass, while for the data we consider here, the pion mass is much larger than the muon mass. 
For more discussion of this point, we refer to Ref.~\cite{Golterman:2017njs}.

%%%%%%%%%%%%%%%%%%%
\begin{figure}[t]
\vspace*{4ex}
\centering
\includegraphics*[width=6cm]{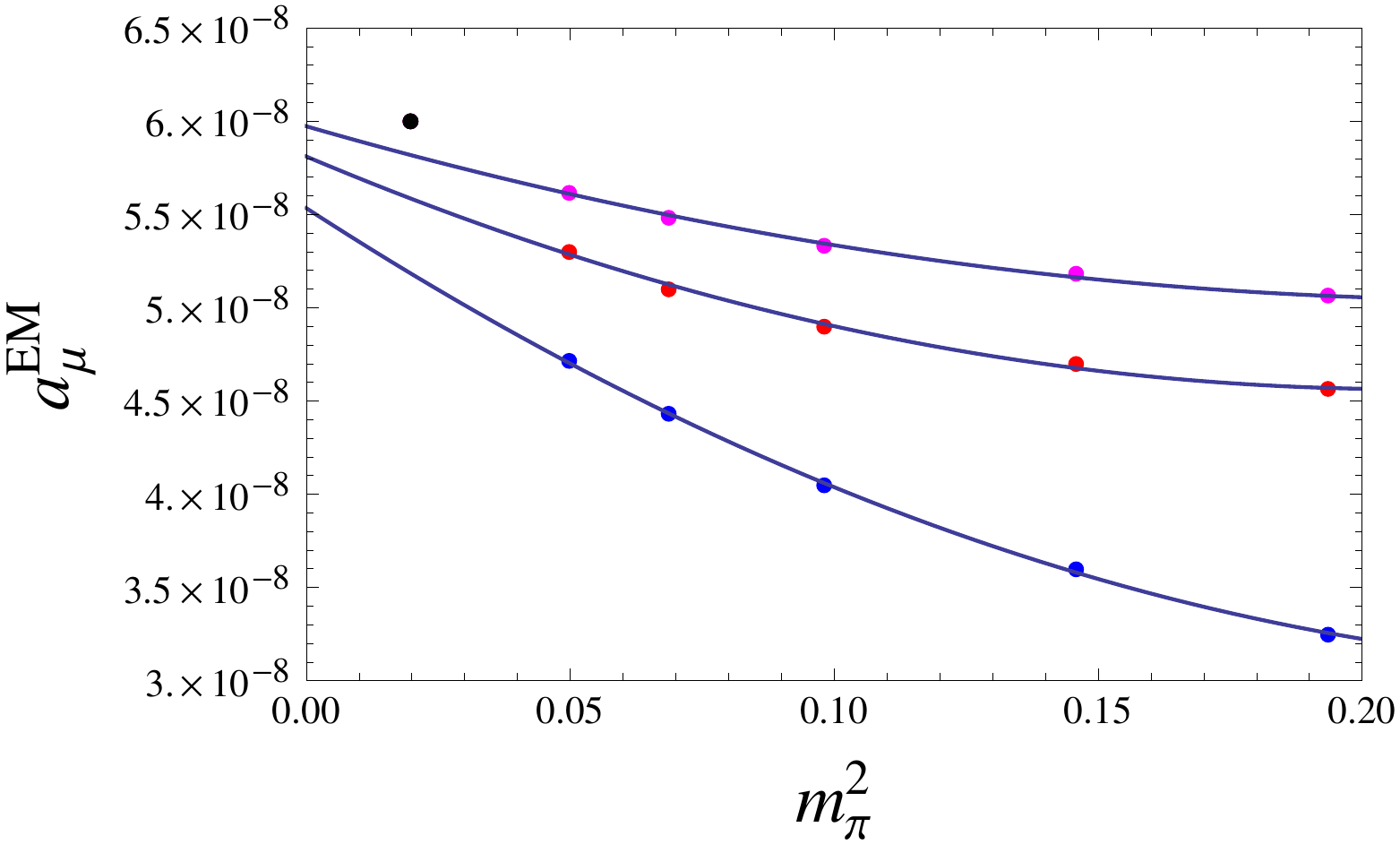}
\hspace{1ex}
\includegraphics*[width=6cm]{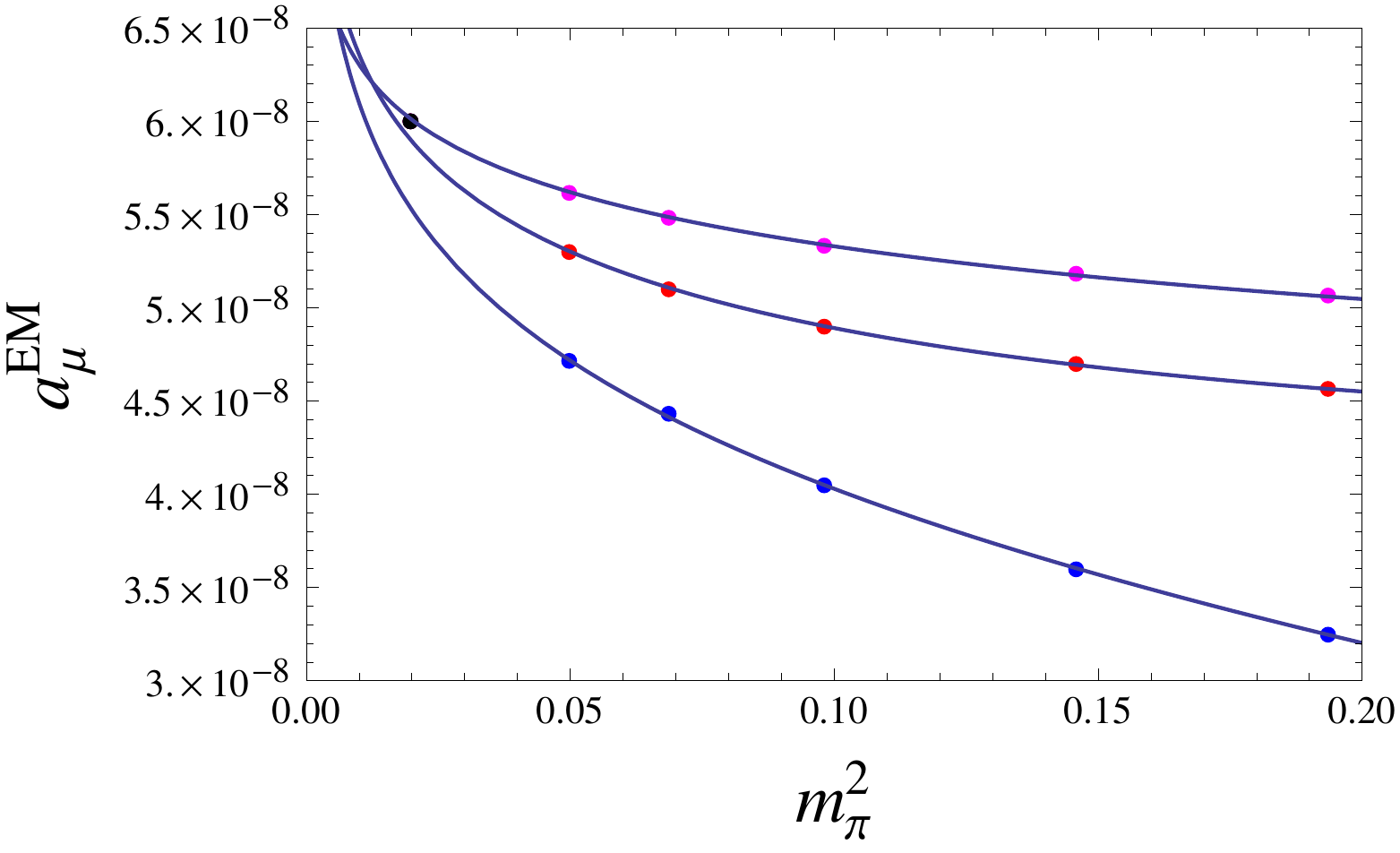}\\
\vspace*{4ex}
\includegraphics*[width=6cm]{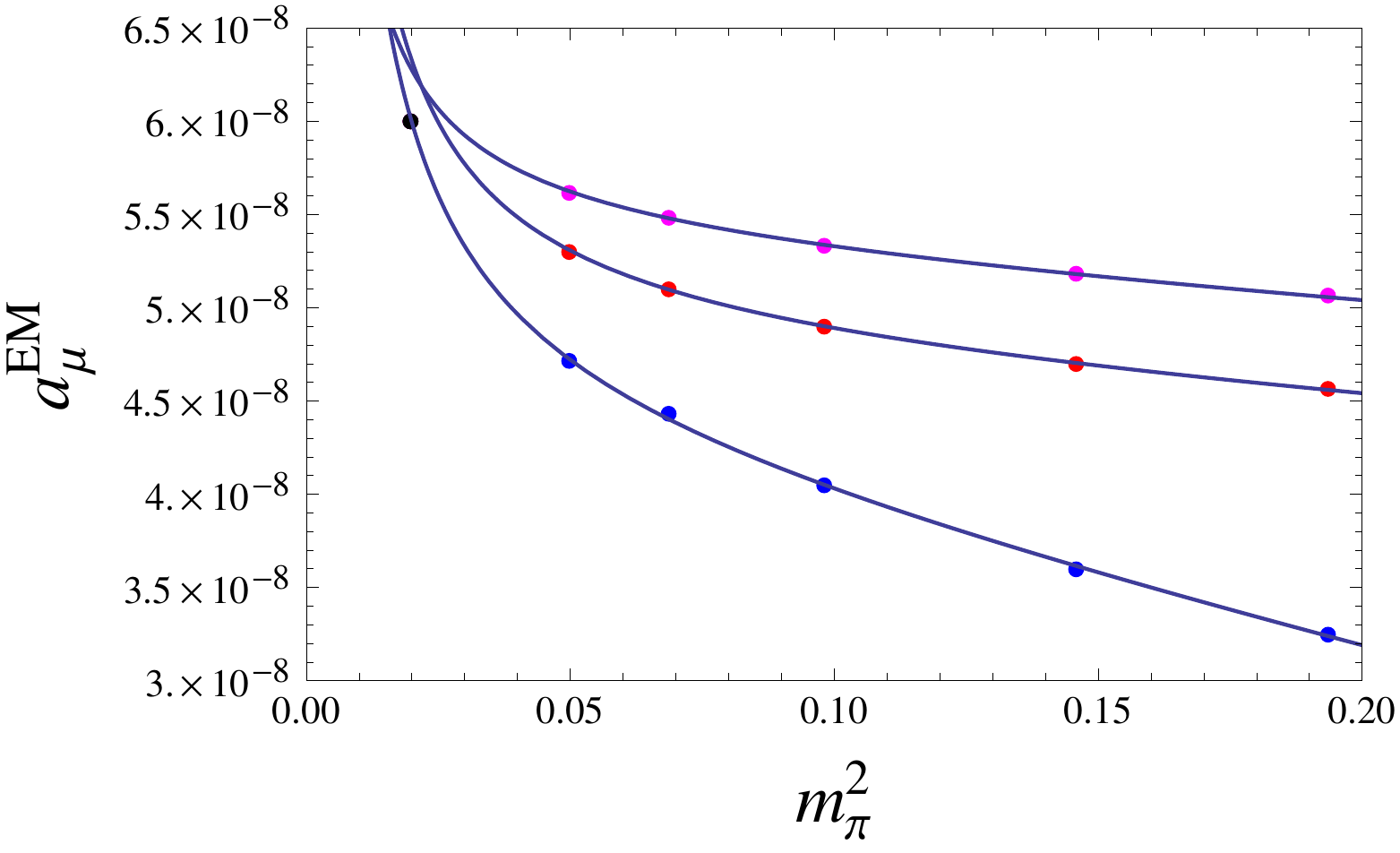}
\caption%
{The unmodified and ETMC- and HPQCD-improved versions of 
$a^{\rm HLO,EM}_\mu(0.1)$ as a function of $m_\pi^2$. In each plot, the upper 
(magenta) data points are HPQCD-improved, the middle (red) data points 
ETMC-improved and the lower (blue) data points unimproved.
The (black) point in the upper left corner of each plot is the 
``physical'' point, $a^{\rm HLO,EM}_\mu(0.1)=6.00\times 10^{-8}$.
Fits are ``quadratic'' (upper left panel), ``log'' (upper right panel)
and ``inverse'' (lower panel). For further explanation, see text.}
\label{fig-3}
\vspace*{-4ex}
\end{figure}
%%%%%%%%%%%%%%%%%%%
We repeated this test for the EM case, and the corresponding plots are shown in 
Fig.~\ref{fig-3}.   Here we lack the direct comparison with experimental
data, but otherwise, the exercise can be carried out in the same manner as for the $I=1$ case.
We thus have to assume that also in this case the representation of the EM HVP by ChPT
to NNLO (augmented with the $C$ term) is good enough up to $Q^2_{max}$ equal to 
$0.1-0.2$~GeV$^2$.
\begin{table}[thb]
  \small
  \centering
  \caption{Values for $a_\mu^{\rm HLO,EM}(0.1)\times 10^8$ for the three types of
fit ({\it cf.} Eq.~(\ref{fits})) and the three data sets. For reference, the correct
model value is $a_\mu^{\rm HLO,EM}(0.1)\times 10^8=6.00$.}
  \label{tab-3}% Give a unique label
  \begin{tabular}{c|c|c|c}\toprule
    & unimproved data & ETMC-improved data  & HPQCD-improved data  \\\midrule
  quadratic fit & 5.19 & 5.59 & 5.82 \\
  logarithmic fit & 5.55 & 5.91 & 6.02 \\
  inverse fit & 6.04 & 6.37 & 6.30 \\  \bottomrule
  \end{tabular}
\end{table}
The numerical results analogous to those of Table~\ref{tab-2} are shown in Table~\ref{tab-3}.  In this case, systematic error estimates obtained
by taking the relative difference between the quadratic and logarithmic fits amounts to
6\%, 5\% and 3\% for the unimproved, ETMC-improved, and HPQCD-improved data.  
While these numbers are slightly better than those for the $I=1$ case, the conclusion is 
the same:   both variants of the ETMC trick do not allow chiral extrapolations with an
accuracy of less than 1\%.   Again, omitting the largest pion mass does not make a 
significant difference.

\section{Conclusion}\label{conclusion} 
%%####%%
We used a ChPT-inspired model to investigate the extrapolation
of the leading-order hadronic contribution to the muon anomalous magnetic 
moment, $a_\mu^{\rm HLO}$, from lattice pion masses of order 200~to~400~MeV 
to the physical pion mass. We found that such pion masses are too large to 
allow for a reliable extrapolation, if the aim is an extrapolation error of 
less than 1\%. This is true even if various tricks to improve the 
extrapolation are employed \cite{Fengetal,HPQCD}.   

In order to perform our study, we had to make certain
assumptions. First, we assumed that useful insight into the pion mass 
dependence could be obtained by focussing on the contribution to 
$a_\mu^{\rm HLO}$ up to $Q^2_{max}=0.1$~GeV$^2$. This restriction 
is necessary if we want to take advantage of information on the
mass dependence from ChPT, since only in this range ChPT 
provides a reasonable representation of the HVP; we believe this
is not a severe restriction. 
Changing $Q^2_{max}$ to $0.2$~GeV$^2$ makes no qualitative difference 
to our conclusions.

Second, we assumed Eq.~(\ref{LECeff}) for the dependence of the
effective LECs $C_{93,{\rm latt}}^{\rm eff}$ and $C_{\rm latt}^{\rm eff}$ on the pion mass. 
While this is a phenomenological assumption, we note that this 
assumption is in accordance with the ideas underlying the ETMC and 
HPQCD tricks, so that those tricks should work well if 
indeed this assumption is correct. There are 
two reasons that the modified extrapolations nevertheless do not work 
well enough to achieve the desired better than 1\% accuracy.  One is the 
fact that in addition to the physics of the $\rho$, the two-pion 
intermediate state contributing to the non-analytic terms in 
Eq.~(\ref{chpt}), also beyond leading order,
plays a significant role as well. 
The second reason is that, although  ChPT provides a simple functional form for the chiral extrapolation of $a_\mu^{\rm HLO}(Q^2_{max}=0.1~\mbox{GeV}^2)$ for pion masses much smaller than the muon mass, this is not useful in practice, so that one needs to rely on phenomenological fit forms, such as those of Eq.~(\ref{fits}).

In order to eliminate the systematic error from the chiral extrapolation, 
which we showed to be very difficult to estimate reliably, one  
needs to compute $a_\mu^{\rm HLO}$ at, or close to, the
physical pion mass.  In contrast to the experience with simpler quantities 
such as meson masses and decay constants, even an extrapolation 
from approximately 200~MeV pions turns out to be a long extrapolation.

It would be interesting to consider the case in which extrapolation 
from larger than physical pion masses is combined with direct computation 
near the physical pion mass in order to reduce the total error 
on the final result. This case falls outside the scope of the study
presented here, because the trade-off between extrapolation and
computation at the physical point is expected to depend on the statistical 
errors associated with the ensembles used for each pion mass. However, our 
results imply that also in this case a careful study should be performed.  The methodology developed in this paper 
can be easily adapted to different pion masses and extended to take 
into account lattice statistics, and we thus recommend such a study for any
computation of the HVP contribution to the muon anomalous magnetic 
moment that relies on extrapolation from larger than physics pion masses.

\vspace{3ex}
%\newpage
\noindent {\bf Acknowledgments}
We would like to thank Christopher Aubin, Tom Blum and Cheng Tu for 
discussions, and Doug Toussaint for providing us with unpublished
hadronic quantities obtained by the MILC collaboration.
This material is based 
upon work supported by the U.S. Department of Energy,  Office of Science,
Office of High Energy Physics, under Award Number DE-FG03-92ER40711
(MG). 
KM is supported by a grant from the Natural Sciences and Engineering Research
Council of Canada.  SP is supported by CICYTFEDER-FPA2014-55613-P, 2014-SGR-1450 and the CERCA Program/Generalitat de Catalunya. 
%\clearpage
\bibliography{lattice2017}

%%%%%%%%%%%%%%%%%%%%%%%%%%%%%%%%%%%%%%%%%%%%%%%%%%%%%%%%%%%%%%%%%%%%%%%%%%%%%
\end{document}